\documentclass[twocolumn]{emulateapj}

\usepackage{natbib}
\usepackage{graphicx}
\usepackage{txfonts}

\newcommand{\sfrunit}{\,{\rm M}_\odot\,{\rm yr}^{-1}}
\newcommand{\lumunit}{\,{\rm erg}\,{\rm s}^{-1}}

\slugcomment{Accepted by ApJ}

\shorttitle{Constraints on the co-evolution of BH and galaxies}
\shortauthors{Zheng et al.}

\begin{document}

\title{Observational constraints on the co-evolution of supermassive black holes and galaxies}


\author{X.\ Z.\ Zheng,\altaffilmark{1} E.\ F.\ Bell,\altaffilmark{2,3} 
R. S.\ Somerville,\altaffilmark{2,4} H.-W. Rix,\altaffilmark{2}
K.\ Jahnke,\altaffilmark{2} F.\ Fontanot,\altaffilmark{2,5}  
G.\ H.\ Rieke,\altaffilmark{6} D.\ Schiminovich,\altaffilmark{7}
K.\ Meisenheimer\altaffilmark{2}} 

\altaffiltext{1}{Purple Mountain Observatory, Chinese Academy of Sciences, West Beijing Road 2, Nanjing 210008, China; xzzheng@pmo.ac.cn}
\altaffiltext{2}{Max-Planck Institut f\"ur Astronomie, K\"onigstuhl 17, D-69117 Heidelberg, Germany} 
\altaffiltext{3}{University of Michigan, 500 Church St., Ann Arbor, MI 48109}
\altaffiltext{4}{Space Telescope Science Institute, 3700 San Martin Drive, Baltimore, MD 21218}
\altaffiltext{5}{INAF-Osservatorio Astronomico, Via Tiepolo 11, I-34131 Trieste, Italy}
\altaffiltext{6}{Steward Observatory, University of Arizona, 933 N Cherry Ave, Tucson, AZ 85721} 
\altaffiltext{7}{Department of Astronomy, Columbia University, New York, NY 10027}

\begin{abstract}
The star formation rate (SFR) and black hole accretion rate (BHAR) 
functions are measured to be proportional to each other at $z \la 3$.  
This close correspondence between SF and BHA would 
naturally yield a BH mass--galaxy mass correlation, 
whereas a BH mass--bulge mass correlation is observed.  
To explore this apparent contradiction we study the SF 
in spheroid-dominated galaxies between $z=1$ and the present day.  
We use 903 galaxies from the COMBO-17 survey with M$_\ast >2\times
10^{10}$\,M$_\odot$, ultraviolet and infrared-derived SFRs 
from Spitzer and GALEX, and 
morphologies from GEMS HST/ACS imaging.  
Using stacking techniques, we find that $<$25\% of all SF occurs in
spheroid-dominated galaxies (S\'ersic index $n>$2.5), while the BHAR
that we would expect if the global scalings held is three times
higher. This rules out the simplest picture of co-evolution, in which
SF and BHA trace each other at all times. These results could be
explained if SF and BHA occur in the same events, but offset in time,
for example at different stages of a merger event. However, one would
then expect to see the corresponding star formation activity in
early-stage mergers, in conflict with observations.  We
conclude that the major episodes of SF and BHA occur in different
events, with the bulk of SF happening in isolated disks and most BHA
occurring in major mergers. The apparent global co-evolution results
from the regulation of the BH growth by the potential well of the
galactic spheroid, which includes a major contribution from disrupted
disk stars.
\end{abstract}


\section{Introduction}

The last decade has seen the discovery and characterization of an
unexpectedly tight correlation between the mass of supermassive black
holes (SMBH; $M_{\rm BH}$) and the mass ($M_{\rm bulge}$) or velocity
dispersion of their host galaxy's bulge
\citep{Magorrian98,Ferrarese00,Gebhardt00,Marconi03,Haring04}.  The
black hole mass appears to be most strongly correlated with the bulge
mass, not the total mass \citep{Kormendy01,Ho07}; the scatter in this
relation between bulge mass and black hole mass is estimated to be
less than a factor of two \citep{Haring04}.

This relationship indicates that galaxy and black hole formation and 
evolution are interconnected.\footnote{\citet{Peng07}, 
noting that multiple generations of mergers would reduce the scatter
of a weak or non-existent $M_{\rm bulge}-M_{\rm BH}$ relation
essentially through the central limit theorem, found that
more than 10 generations of major mergers would be required to 
imprint a tight $M_{\rm bulge}-M_{\rm BH}$ relation in a population 
that was initially uncorrelated.  Such a large number of major mergers
is highly unlikely, arguing that while mergers may tighten
this relation much of the correlation must be imprinted
through a strong interconnection between $M_{\rm bulge}$ and $M_{\rm 
BH}$.}
In its weak form, such an interconnection could result if the black hole 
growth is limited by the wider galaxy environment (`dog wagging the tail'). 
Stronger forms of interconnection are also possible. 
The energy released by a growing SMBH is sufficient, if it couples 
effectively with its surroundings, to have dramatic consequences 
on the galaxy (`tail wagging the dog'). For example, the injection 
of energy into hot halo gas in galaxy clusters is indicated by 
the inflation of bubbles and the propagation of sound waves 
\citep{McNamara00,McNamara05,Fabian03,Rafferty06}.
 Furthermore, it is possible that SMBHs with high accretion rates 
drive powerful outflows that remove (at least some) cold gas 
from galaxies \citep{Chartas03,Crenshaw03,Pounds03,Tremonti07}.
In fact, feedback from an accreting SMBH/AGN may help address 
many of the most difficult issues affecting models of galaxy 
evolution in a cosmological contest: e.g., overcooling in massive 
halos and late star formation in elliptical galaxies 
\citep{Binney95,Croton06,Bower06,Monaco07,Somerville08}.

A diversity of models has been constructed to explore the
interrelationship between bulge mass and black hole mass, featuring
both weak and strong aspects of the possible interconnections
\citep[][and references therein]{Somerville08}.  However, the details
of the physical processes governing BH formation and growth, and the
physical mechanisms whereby SMBHs couple with their host galaxies and
surroundings remain poorly understood.  It will be impossible to model
the full range of interconnected processes from theoretical first
principles for some time, as the physics of the formation of SMBHs and
galaxy evolution spans at least $\simeq$ 7 orders of magnitude in
scale. Empirical and phenomenological constraints are therefore
extremely important.

A common feature of most ``unified'' models of SMBH and galaxy
formation is the concept of co-evolution, whereby the SMBHs and their
galaxies evolve under each other's influence. The term co-evolution
means many things to many people
\citep[e.g.,][]{Silk98,Kauffmann00,Wyithe03,Granato04,Merloni04,Haiman04,Cattaneo05,Fontanot06}. For
the purpose of having strawmen to take aim at, we outline three
possible scenarios:

\begin{itemize}
\item {\bf Strong Co-evolution: } In its strongest sense, co-evolution
  implies that bulges and black holes grow {\it together}, i.e., in
  the same objects and at the same time. In this case, BH growth could
  be thought of as a sort of a `tax': whenever star formation occurs,
  the SMBH accretes some fraction of the mass involved. This scenario
  in its simplest form can already be ruled out --- there are many
  examples of star-forming bulges without AGN activity and bulges with
  AGN activity and no SF (e.g., M87).
\item {\bf Time Offset: } One could imagine a scenario in which bulge
  stars and SMBHs are formed in the same event, but on different
  timescales and/or stages of the event. For example, star formation
  might occur predominantly in the early stages of a merger, while BH
  growth might occur in the later stages
  \citep[e.g.,][]{DiMatteo05,Hopkins05,Hopkins06}. Such a picture
  would predict little evolution in the BH-bulge mass correlation for
  an unbiased ensemble, but galaxies in the throes of their bulge
  building events (or black holes in their most rapid growth phase)
  could exhibit significant deviations from the scaling relations;
  this hypothesis is considerably more challenging to rule out.
\item {\bf Regulated Growth: } A third possibility is that the major
  episodes of SF and BH accretion occur in different events, but the
  growth of one component regulates that of the other. For example, if
  black hole growth is regulated by feedback from the AGN itself, as
  suggested by many workers
  \citep[e.g.,][]{Silk98,Murray05,DiMatteo05}, the potential well of
  the pre-existing bulge determines how much energy is needed to stop
  further accretion and halt the growth of the black hole, and hence
  the final black hole mass. It is possible, even likely, that the
  bulge stars that dominate this potential well are formed mostly in
  previous star formation ``events'', i.e. in progenitor disks which
  have subsequently merged. In this picture, one would expect a larger
  possible disconnect between observable episodes of SF vs. BH
  accretion {\it activity}, even while the endpoints of this activity
  (bulges and black holes) are expected to end up being tightly
  related.
\end{itemize}

There are a number of ways to attack the problem of co-evolution
observationally.  Obviously, measuring the BH-bulge mass correlation
at the present day at extremely low or high mass will provide powerful
insights into the processes linking galaxy and SMBH evolution
\citep{Barth05,Lauer07}.  The study of the BH-bulge mass correlation
during BH or bulge-building events is also an important avenue of
approach
\citep[e.g.,][]{Barth05,Greene06,Treu04,Treu07,Woo06,Peng06,Borys05}.
The most commonly-used technique in the latter approach is studying
the masses of actively-accreting Type I AGN for which BH masses can be
estimated using an empirically-calibrated combination of luminosity
and broad line region line width \citep[][and references
  therein]{Wandel99,Baskin05,Kaspi05,Vestergaard06}; current results
are controversial but tentatively support larger BH to bulge mass
ratios at early times than those seen today \citep[][see also, e.g.,
  \citealt{Alexander05} for the smaller ratios of black hole to bulge
  mass in rapidly-star-forming galaxies detected in the
  submm]{Treu04,Woo06,Peng06,Shields06}.

With current generations of wide and deep cosmological surveys,
another means has opened up: comparison of evolution of the
BH accretion rate (BHAR) and that of the SFR of galaxies 
\citep[\citealt{Silverman09}; see \citealt{Boyle98} and][for early efforts]{Franceschini99}. 
The history of BH accretion has been explored using
primarily optical and X-ray techniques, with important insights being
gained from observations in the radio and infrared
\citep{Miyaji00,Ueda03,Hasinger05,LaFranca05,Barger05,Donley05,Brown06,Richards06a}.  
Furthermore, a decade of intensive multi-wavelength and redshift
surveying has placed powerful and interesting constraints on the
volume-averaged SF history of the Universe
\citep{HB06,PerezGonzalez05,LeFloch05,Schiminovich05}.
In both cases, the integrated BH accretion and SF histories
give values that match to within a factor of 2 or 3 of the present-day 
BH mass and stellar mass\footnote{In the case of stellar mass density,
recycling of nearly 1/2 of the initial stellar mass back into
interstellar gas must be accounted for to avoid dramatically
overproducing the present-day stellar mass density.} 
densities \citep{Yu02,Marconi04,Borch06,Fardal07,
Wilkins08}; we will discuss in detail
uncertainties and possible inconsistencies
between the integral of the SFR and the present-day stellar
mass in \S \ref{sec:overall}.

The observational maturity of current datasets at $z<1$ allows the use
of a novel angle of attack on the problem of co-evolution.  Recent
multi-wavelength/HST surveys have covered enough volume that SFR
density and BHAR density can be estimated in well-defined subsamples
of galaxies for the first time, offering potentially decisive insight
into not only the statistical evolution of these quantities, but into
whether black holes and bulges are actually growing {\it in the same
  objects and at the same time}.  This is the approach we will adopt
in this paper.  To do this, we first examine the global SFR and BHAR
(\S \ref{sec:overall}), then the multiplicity functions\footnote{The
  comoving number density of sources with a given SFR or BHAR ---
  e.g., a luminosity function is a multiplicity function} of SFR and
BHAR as a function of redshift (\S \ref{sec:multifuncs}).  In \S
\ref{sec:host}, we present new measurements of SF activity in mass and
morphology-limited samples of galaxies, and examine accretion activity
driven by massive BHs in massive spheroids under assumptions widely
used in literature.  In \S \ref{sec:summ}, we discuss the results in
the context of co-evolution and summarize our conclusions.  Throughout
the paper we assume a cosmology with
H$_0$\,=\,70\,km\,s$^{-1}$\,Mpc$^{-1}$, $\Omega_{\rm M}$\,=\,0.3 and
$\Omega_{\Lambda}$\,=\,0.7.

\section{The cosmic star formation and BH accretion histories} \label{sec:overall}

We first attempt to give some context to the analysis that 
will follow by exploring the cosmic star formation 
and black hole accretion histories.

\subsection{The cosmic star formation history}\label{sfrhis}

\begin{figure*}[t] \centering
  \includegraphics[width=0.80\textwidth]{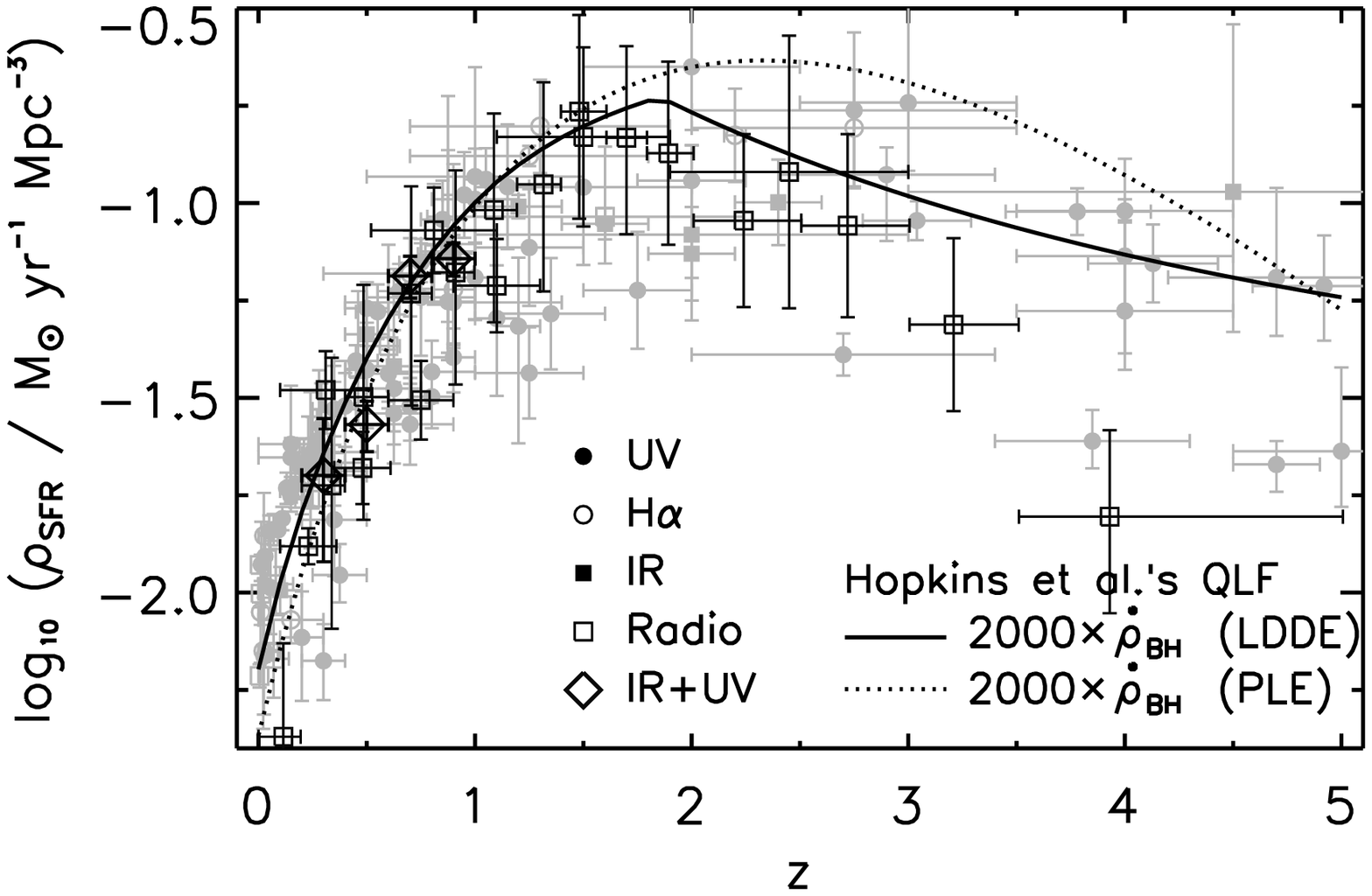}
  \includegraphics[width=0.80\textwidth]{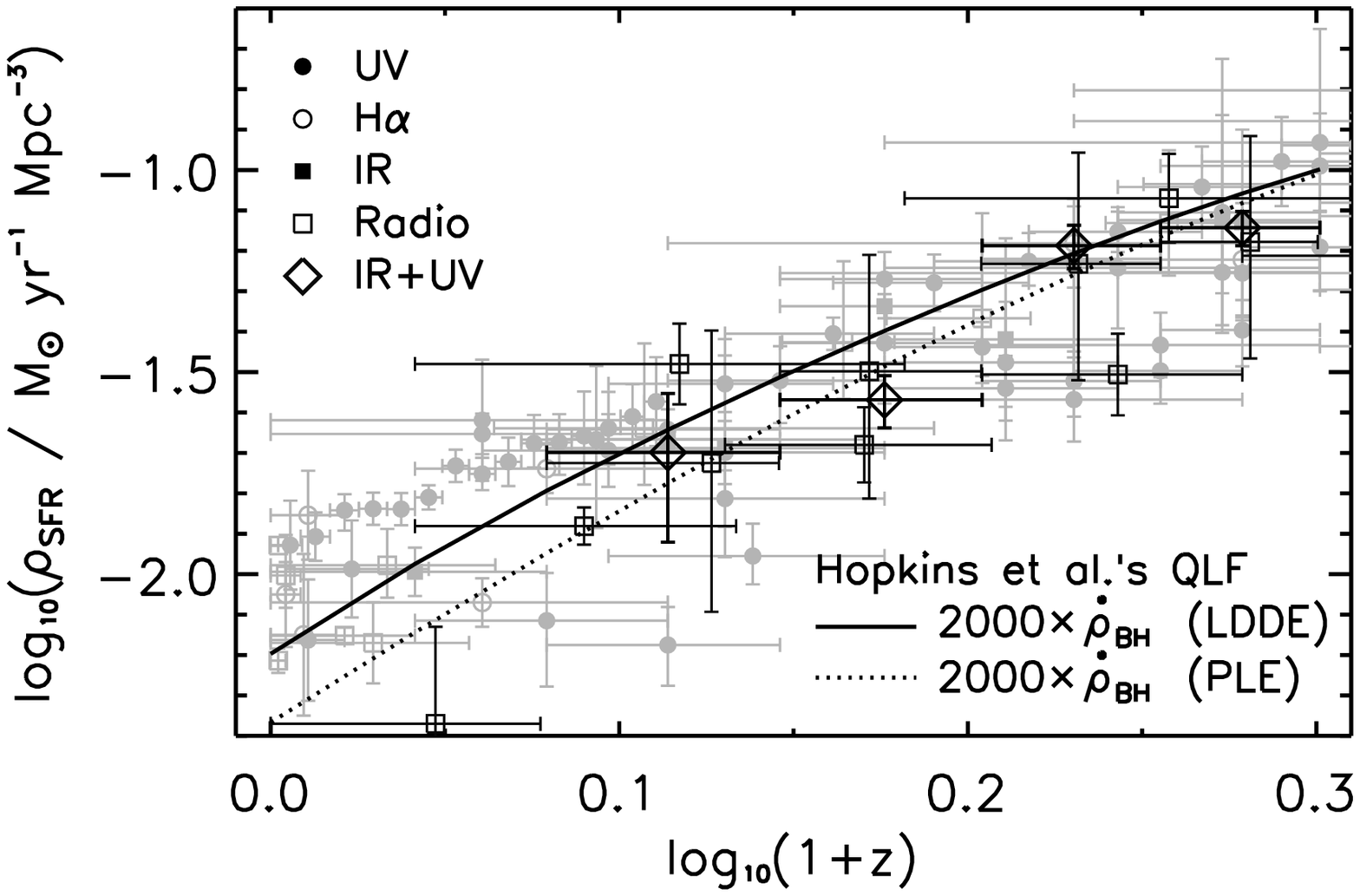}
\caption{{\it Top:} Comparison between the volume-averaged SF
  history and the volume-averaged BH accretion history; 
{\it Bottom:} The same comparison but only in the redshift range $0<z<1$. 
The {\it gray} data points come from the compilation of available measurements 
by \citet{HB06}.
The {\it black squares} are radio measurements from \citet{Seymour08}, \citet{Smolcic09}
and \citet{Dunne08}. The {\it black diamonds} represent IR+UV measurements from
\citet{Bell07}. All measurements are converted to a Kroupa IMF and the same cosmology
adopted here. 
The BH accretion history is derived from AGN bolometric luminosity functions
given in HRH07 by assuming a radiation efficiency $\epsilon = 0.1$.
The BH accretion history is shifted upwards by 3.3\,dex (a factor of 2000).}
\label{sfr}
\end{figure*}

During the last decade, a number of studies have contributed to
determining the average SFR per co-moving volume at different cosmic
epochs with various SFR estimators (e.g., UV, H$\alpha$, IR and
radio). The available measurements from the literature compiled by
\citet{HB06}, and complemented with recent measurements from \citet{Seymour08},
\citet{Smolcic09} and \citet{Dunne08} are shown in Figure~\ref{sfr}.
We include also recent estimates from the use of UV and 24{\micron}
data to account for both unobscured and obscured star formation 
at $z<1$ \citep{Zheng07,Bell07}.  Here all
measurements are corrected to our adopted cosmology and a
\citet{Kroupa01} initial mass function (IMF). 
It can be seen from Figure~\ref{sfr} that the
cosmic SFR density increases rapidly with redshift, peaks at $z\sim
1.5$, then becomes flat or declines at $z > 1.5$, although the
uncertainties are large at these early epochs.  
The cosmic SFR density decreases by an order of magnitude from the
$z=1$ to the present day, following $\rho_{\rm SFR} \sim
(1+z)^{2.9\pm0.2}$. 

A complementary measurement to the cosmic SFR density is the cosmic
stellar mass density. The assembly of stellar mass should be
consistent with the integral of the cosmic SFR density.  Current
measurements show a reasonably good agreement between the two, at
least at redshifts less than about unity \citep{Borch06,Bell07,Wilkins08};
about 40-50\% of local stars were formed
since $z=1$ \citep{Dickinson03,Fontana03,Drory05,Borch06,Rudnick06}.
As an aside, it is worth noting that 
there is a possible mismatch between 
the integral of the cosmic SFR density 
at $z>1$ with the stellar mass observed to be in place at 
$z=1$: it appears that the integral is a factor of $\sim 3$
in excess of the stellar mass formed by $z=1$.  The origin 
of this mismatch is currently not well-understood.
It is possible that this discrepancy signals a break-down in the utility
of SFR indicators or a change or break-down of a
universally-applicable stellar IMF.  It should also be
noted that at $z>1$ the estimates of SFR (and stellar mass) 
are highly uncertain; in particular, the relatively sensitive
UV measurements need to be corrected for 
substantial (factors of a few) dust extinction using defensible but 
uncertain
recipes \citep{Reddy08}, and measurements
of the rest-frame IR and radio probe only the most luminous
systems (see, e.g., \citealt{Dunne08} for substantial progress
towards this goal using stacking of radio data; it is interesting
that their resulting SFR densities are much lower than those reported
previously at $z \ge 1.5$).  Yet, 
for our purposes, the possibility of a mismatch between 
SFR and stellar mass at $z>1$ is of relatively little importance; 
the focus of our work is at the better-constrained $z<1$ redshift range.

\subsection{The cosmic BH accretion history}
\label{bhacchis}

Observed (optically ``bright'') BH accretion apparently can account
for nearly all of the BH mass seen in remnants today, indicating that
the bulk of BH mass growth occurs during a luminous AGN phase with a
radiation efficiency $\epsilon$ $\sim 0.1$
\citep[e.g.,][]{Yu02,Marconi04,Shankar04}. The observed AGN luminosity
function is therefore a reasonable probe of the cosmic BH accretion history.
>From deep cosmological surveys performed with modern observational
facilities, AGN luminosity functions have been determined out to $z\sim 4$ 
in hard X-ray \citep[e.g.,][]{Ueda03,LaFranca05,Barger05}, soft X-ray
\citep[e.g.,][]{Miyaji00,Hasinger05},
optical \citep[e.g.,][]{Croom04,Richards06b,Fontanot07}
and mid-IR bands \citep[e.g.,][]{Brown06,Richards06a}.
The intrinsic spectral energy distribution (SED) of AGN is thought to
be only a function of luminosity \citep[e.g.,][]{Marconi04}; and
accounting for obscuration, the AGN luminosity functions of different
bands are correlated through the SED.  \citet[][hereafter 
HRH07]{Hopkins07}
presented bolometric AGN LFs derived from the combination of AGN LFs
measured in multi-wavelength bands, from mid-IR through hard X-ray
(see their paper and references therein for details of the AGN LFs and
related uncertainties).   We
caution that Compton-thick AGN are not counted in the HRH07 bolometric
LFs. Correction for the missing accretion is suggested to be $\leq
0.15$\,dex (a factor of 1.4; see HRH07 for more
discussion).
HRH07 presented several evolution models which fit the AGN
bolometric luminosity function at different cosmic epochs.  We adopted
the best-fit pure luminosity evolution (PLE) model and the
luminosity-dependent density evolution (LDDE) model of the bolometric
LF $\Phi (L_{\rm bol},z)$ from HRH07 to calculate the BHAR
per co-moving volume as follows:
\begin{equation} \label{eq:acc}
 \dot{\rho}_{\rm BH}(z) = \int^\infty_0\frac{(1-\epsilon)\,L_{\rm 
bol}}{\epsilon\, c^2} \Phi (L_{\rm bol},z)\,{\rm d}L_{\rm bol}, 
\end{equation}
where $L_{\rm bol}$ is the bolometric luminosity and we have adopted
$\epsilon = 0.1$. By default $L_{\rm bol}$ is given in units of
$\lumunit$.  We integrate the bolometric luminosity over the range
$43<\log L_{\rm bol}<49$.
%
%
The cosmic BHAR density peaks at $z\sim 1.9$, and dramatically
decreases to the present day.  At $z>1.9$, the cosmic BHAR density is
likely to decline or stay flat to earlier cosmic times although the
uncertainties are large \citep[see also][]{Barger05}.  Roughly
speaking, the cosmic BHAR density increases as $\dot{\rho}_{\rm BH}
\sim (1+z)^{\alpha}$ and $\alpha\sim 3.4$ up to $z=1$.  Considering
the uncertainties of different band LFs and uncertainties in
bolometric correction, a typical error of $\sim$0.6 is adopted for the
power index $\alpha$.

\subsection{Comparison between the cosmic star formation and BH accretion
  histories} \label{comsfrbh}

Figure~\ref{sfr} shows the cosmic SFR density to be in good agreement
with the cosmic BHAR density scaled by a factor of 2000 over a wide
redshift range from $z\sim 0$ to $z\sim 5$.  The factor 2000 is the
ratio of the integral of the cosmic SFR density to the integral of the
cosmic BHAR density within the redshift range from $z=0$ to $z=1$,
where both quantities are well determined.  In the redshift range
$0<z<1$ a quantitative comparison between the cosmic SFR density and
BHAR density can be made: they increase with redshift with almost the
same slope of $\dot{\rho} \sim$ (1+z)$^{3}$ within the
errors.  Figure~\ref{sfr} suggests that the cosmic BH accretion
history is parallel to the cosmic SF history.

If the strong co-evolution hypothesis is correct, the shift between
the volume-averaged SFR and volume-averaged BHAR (a factor of 2000;
3.3\,dex) should be consistent with the difference between the local
stellar mass density and BH mass density.  The integral of the SFR
needs to be corrected for the recycling of mass back into gas during 
the process of stellar evolution; with our adopted 
IMF, almost 50\% of the initially-formed 
stellar mass is returned to the ISM within 5--7\,Gyr.
Thus, one expects a ratio of local stellar mass density to 
BH mass density of $\sim 1000$.
This is consistent with the measurements of local
stellar mass density $\rho_{\ast,0} = 3.0^{+0.8}_{-0.6} \times
10^8$\,M$_\odot$\,Mpc$^{-3}$ \citep[e.g.,][]{Bell03} and local BH mass density
$\rho_{\rm BH,0} = 4.6^{+1.9}_{-1.4} \times
10^5$\,M$_\odot$\,Mpc$^{-3}$ \citep[e.g.,][]{Marconi04}.

Thus, somewhat remarkably, it appears that the histories
of SMBH accretion and star formation are very similar 
in shape, offset by a ratio consistent with the ratio between 
BH mass and stellar mass density at the present day.  
Put differently, it appears that the ratio of stellar mass
to SMBH mass density is {\it independent} of redshift, in 
seeming accord with the co-evolution picture.

\begin{deluxetable}{cccc}
\tabletypesize{}
\tablewidth{0pt}
\tablecaption{Best-fit parameters for the SFR functions.$^\mathrm{a}$}
\tablehead{ & & $\log SFR^\ast$ & $\log \Phi^\ast$  \\
           $z_{\rm start}$ & $z_{\rm end}$ & ($M_\sun$\,yr$^{-1}$) & (Mpc$^{-3}$\,dex$^{-1}$) }
\startdata
0.2 & 0.4 & 1.09 & -2.80  \\
0.4 & 0.6 & 1.11 & -2.77  \\
0.6 & 0.8 & 1.12 & -2.63  \\
0.8 & 1.0 & 1.15 & -2.90  \\
\enddata
\tablenotetext{a}{The SFR functions are given in the form of ``double power-law'', $\log \Phi = \log \Phi^\ast+\alpha\,\log SFR$, with $\alpha = -0.6$ and $\alpha = -2.2$ for $SFR$ below and above $SFR^\ast$, respectively.
}
\label{lfp}
\end{deluxetable}

\begin{figure*} \centering
  \includegraphics[width=0.80\textwidth]{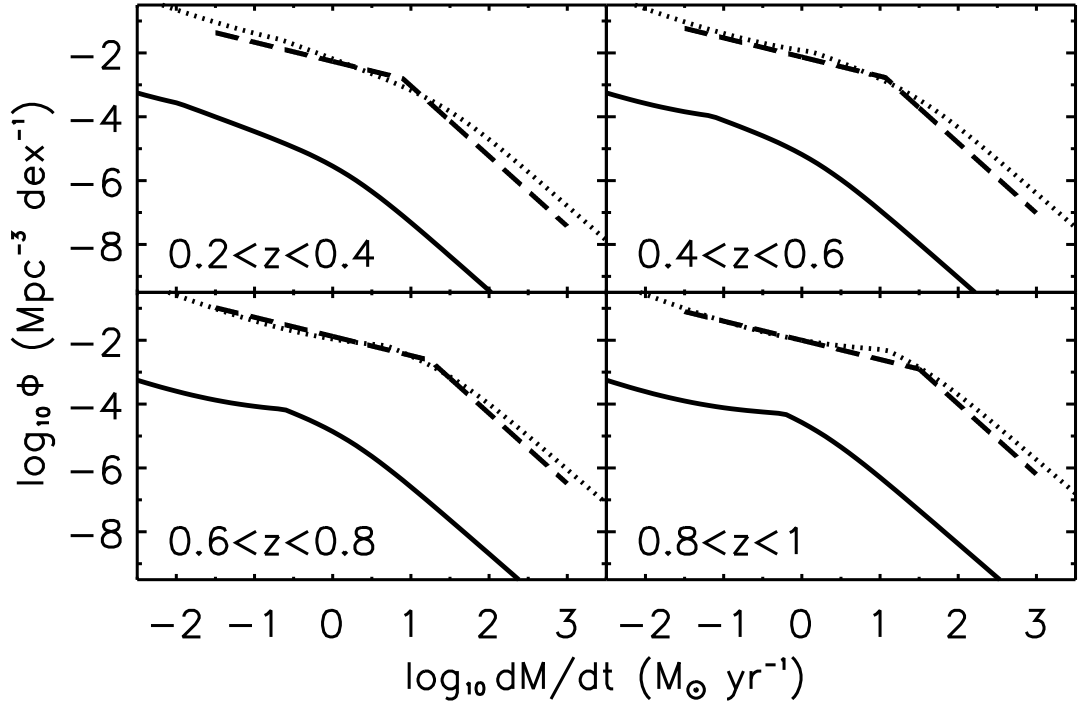}
\caption{Comparison between the SFR functions (the {\it dashed} lines)
and the BHAR functions (the {\it solid} lines) in four redshift bins 
from $z=0.2$ to $z=1$.
The SFR functions are obtained by fitting the data points in \citet{Bell07} with
``double power-law'' function from \citet[][see text for details]{Sanders03}.
The BHAR function is the conversion of the AGN bolometric luminosity function 
 from HRH07 with a radiation efficiency $\epsilon = 0.1$. 
The {\it dotted} lines are the BHAR functions shifted by 20 along the x-axis and 
100 along the y-axis.}
\label{sfrfuncs}
\end{figure*}

\section{Statistical links between SF and BH accretion events}\label{sec:multifuncs}

The overall BH accretion/SF mass ratio results from the sum of 
individual BH accretion/SF events in the same cosmic epoch.  
Here, a BH accretion ``event'' means
a SMBH in the active phase (i.e., AGN) and a SF event refers to a
galaxy of a given SFR.  In this section we address whether the
intensity of star formation events is statistically correlated with
the intensity of BH accretion events.

Following the description in \S \ref{bhacchis}, we
convert the AGN bolometric luminosity functions of HRH07 into BHAR
functions by assuming $\epsilon = 0.1$. We take the BHAR functions
described by a double power-law as describing the statistics of BH
accretion events.
\citet{Bell07} estimated the SFR using bolometric (UV+IR) luminosity 
for a sample of 7506 galaxies and derived SFR functions in four
redshift slices between $z=0.2$ and $z=1$.  Motivated by local 
IR luminosity functions that are well-fit with a ``double power-law'' 
shape \protect\citep[having a form of $\Phi (L) \propto L^\alpha$ 
with $\alpha = -0.6 (\pm 0.1)$ and 
 $\alpha = -2.2 (\pm 0.1)$ for $L< L^\ast$ and $L> L^\ast$, 
respectively]{Sanders03}, we re-fit the SFR functions of 
\citet{Bell07} with a double power-law.
The best-fit parameters for these SFR functions are given in Table~\ref{lfp}. 
We adopt these SFR multiplicity functions to describe the
statistics of SF events. 

\begin{figure*} \centering
  \includegraphics[width=0.80\textwidth]{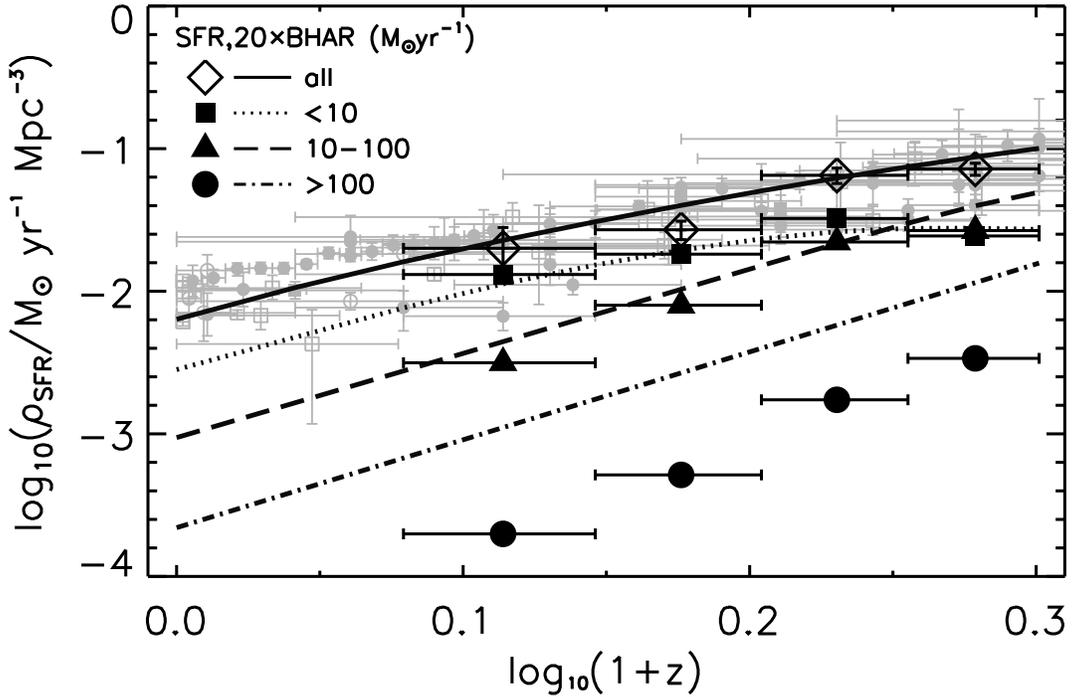}
\caption{The cosmic SFR density ({\it solid symbols}) and BHAR density
({\it lines}) split into high-intensity (circles; the {\it dot-dashed}
line), medium-intensity (triangles; the {\it dashed} lines) and
low-intensity episodes (squares; the {\it dotted} lines).  }
\label{sfrbhar}
\end{figure*}

Figure~\ref{sfrfuncs} shows the adopted BHAR functions and SFR functions.
We have shown in \S \ref{comsfrbh} that the cosmic BH accretion
history tracks the cosmic SF history through a universal scaling
factor of 2000.  Comparing the BHAR function with the SFR function, we
find that the former always tracks the latter in all four redshift
slices from $z=0.2$ to $z=1$ after being re-scaled by a factor of 20
in BHAR and by 100 in number density, as shown in Figure~\ref{sfrfuncs}.
The two scaling factors 20 and
100 are empirically determined to make the agreement between the two
functions as good as possible, particularly in the wide regime around
the SFR function ``knee'' (i.e., 0$<\log (SFR/\sfrunit)<$2), where the
vast majority of the total SF occurs.  
We split SF events by their
intensity into three classes: high-intensity ($SFR>100$\,$\sfrunit$;
i.e., ultraluminous IR galaxies), medium-intensity
($10<SFR<100$\,$\sfrunit$; i.e., luminous IR galaxies) and
low-intensity ($SFR<10$\,$\sfrunit$; i.e., ``normal'' galaxies).
Similarly, we split BH accretion events into the same three classes by
replacing $SFR$ with 20$\times BHAR$.  The three classes of AGN
roughly correspond to luminous quasars, quasars and Seyferts/low-luminosity AGN,
respectively.  From the adopted SFR/BHAR functions, we calculate the
volume-averaged SFRs/BHARs contributed by the three classes,
respectively.  The results are shown in Figure~\ref{sfrbhar}.

As can be seen from Figure~\ref{sfrbhar}, the volume-averaged SFR/BHAR
contained in high-intensity star formation or BH accretion events
decreases dramatically from $z=1$ to $z=0.2$, whereas the decrease of
the volume-averaged SFR/BHAR with decreasing redshift is gradually
slower for the medium-intensity and low-intensity SF and BH accretion
events. The low-intensity SF and BH accretion events dominate the
cosmic SFR/BHAR at low-$z$ and the high-intensity star formation/BH
accretion events start to dominate the cosmic SFR/BHAR at $z>0.8$.  It
is clear that for either the medium-intensity or the low-intensity
class, the volume-averaged SFR matches the volume-averaged BHAR
remarkably well over the redshift range $0.2<z<1$.  For the
high-intensity class, the agreement is poor, possibly because of 
the large uncertainties in the SFR and BHAR functions in this range
($SFR>100$\,$\sfrunit$ or $20\times BHAR > 100$\,$\sfrunit$).  
We conclude that
SF events statistically track BH accretion events over $0.2<z<1$ when
split by intensity.

\section{Characterizing the host galaxies of star formation and BH accretion}\label{sec:host}

In \S 2 and \S 3, we presented evidence that both the distribution 
(multiplicity functions) and integrals of the SF rate and supermassive 
black hole accretion rate evolve similarly at all redshifts, with an offset
in integrated rate of a factor of $\sim 2000$ at all $z \la 1$.  
Taken at face value, such a coincidence may lead one to predict a correlation 
between galaxy mass and supermassive black hole mass, offset by 
a factor of $\sim 1000$ in zero point (where the expected offset is 
a factor of 1000 rather than 2000 because of the recycling of 
stellar mass during the course of stellar evolution).
Such a correlation is not observed; instead, it appears that {\it bulge}
mass and supermassive black hole mass correlate \citep{Kormendy01}.

In this section, we present new measurements of the relationship between SFR 
and BHAR for spheroid-dominated galaxies, in order to investigate the seeming
disconnect between a BH-bulge mass relation on one hand, and the 
close correspondence between SFR and BHAR statistics on the other.

\subsection{Links between SFR, stellar mass and morphology} \label{sfrmass}

\subsubsection{The data}

We address this issue using a deep, wide-area multi-wavelength dataset 
from the extended Chandra Deep Field South (ECDFS).
We use optical photometry, photometric redshift catalogs ($\delta
z/(1+z)\sim 0.02$; \citealt{Wolf04}) and stellar mass estimates
\citep{Borch06} from the COMBO-17 survey for $\sim 9000$ galaxies with
aperture magnitudes $m_{\rm R} < 23.5$ mag and $z<1.1$ in the
$30\farcm 5\times30\arcmin$ ECDFS.  High-resolution (0$\farcs 07$)
HST imaging from the Galaxy Evolution from Morphology and SEDs (GEMS)
survey \citep{Rix04} covers around 800 square arcminutes of the
ECDFS in the F606W and F850LP passbands, 
providing optical morphologies for $\sim$8000 galaxies.  A
two-dimensional light distribution analysis is performed on the F850LP 
imaging data using the
software tool GALFIT, providing S\'ersic index $n$ as a broad measure of the
concentration of a galaxy \citep{Haussler07}. 
\footnote{Surface brightness dimming is likely to
introduce a redshift-dependent systematic effect.
Disk components (typically of relatively lower-surface brightness)
will probably be less frequently detected at higher redshifts
\citep[see][for a related discussion]{Shi09}. A tendency towards
a higher fraction of disk contamination in higher-redshift spheroids
could result, affecting the inferred redshift evolution of the SFR density 
in $n>2.5$ systems.}

Deep far-ultraviolet (FUV; 1350-1750\AA) and near-ultraviolet (NUV;
1750-2800\AA) images centered on the ECDFS were obtained by the {\it
Galaxy Evolution Explorer} ({\it GALEX}: \citealt{Martin05}), with a
field of view of a square degree. The FUV and NUV images have a
typical resolution $\sim5\arcsec$, and a depth of 3.63\,$\mu$Jy at the
5\,$\sigma$ detection level. Data reduction and source detection is
described in \citet{Morrissey05}.  Deep 24\,$\micron$ data of the
ECDFS were taken as part of the MIPS GTO observations
\citep{Rieke04}. A mosaic image was produced with a rectangular field
of 90$\arcmin\times30\arcmin$. The 24\,$\micron$ image has a point
spread function (PSF) with full width at half maximum (FWHM) $\simeq
6\arcsec$. Sources are detected down to 83\,$\mu$Jy at the 5\,$\sigma$
level \citep[see][for details of data reduction, source detection and
photometry]{Papovich04}.

\subsubsection{Sample selection}

The goal of \S 4 is to study the SFR in massive galaxies split by morphology
(or, more precisely, structure, as we use S\'ersic index $n$ to differentiate
between spheroid- and disk-dominated galaxies).  Accordingly, in 
this section, we define the mass-limited, SFR-limited and structurally
(morphologically)-selected galaxy samples that we use for further study.

A sample of 903 massive galaxies ($M_\ast \ge
2\times10^{10}$\,M$_\odot$) in the redshift range of 0.2$<z<$1 is
selected from the COMBO-17/GEMS survey in the ECDFS.  These objects
are located in the overlap area ($\sim$600 square arcminutes) of the
{\it HST, GALEX} and MIPS observations that allow for SFR
measurements.  We have removed X-ray point sources detected in the
Chandra 250\,ks observation \citep{Lehmer05} in order to exclude
AGN-heated dust and AGN UV emission (we describe later how this 
selection criterion is of only modest importance; less than 15\%
of star formation is in such systems).
For the present comparison, we identify
spheroid-dominated galaxies with galaxies of S\'ersic index
$n>2.5$. The criterion $n=2.5$ is a reasonable and reproducible 
separation between concentrated
early-type galaxies and late-type galaxies \citep[][see, e.g., \citealt{vanderWel08} for a discussion of the relationship between morphology and S\'ersic index in the SDSS]{Bell04,Pannella06}. 
In what follows, we use this 903-galaxy sample to define three galaxy 
subsamples: 1) mass limited ($M_\ast \ge 2\times10^{10}$\,M$_\odot$; 903 galaxies); 2)
mass and SFR limited ($SFR\ge 3\sfrunit$; 308 galaxies, SFRs are defined in \S 4.1.3); and 3) mass and
morphology-limited (S\'ersic index $n>2.5$; 493 galaxies).

\subsubsection{Estimating SFR from the UV and IR}\label{essfr}

We divide the galaxies of each subsample into four even ($\Delta z$ = 0.2)
redshift slices. Average UV and IR luminosities are estimated for
galaxies in each redshift bin and each sample.  We use stacking
techniques to estimate the average fluxes of individually undetected
objects in a subset of galaxies (see \citealt{Zheng06} for details
about the 24\,$\micron$ stacking).  This is important for obtaining a
complete SFR estimate for massive spheroid-dominated galaxies because
most such galaxies are intrinsically faint in the UV
and mid-IR.  The stacked fluxes and individually detected fluxes are
combined to obtain the mean flux for the given subset of galaxies. This
procedure was applied to MIPS 24\,$\micron$ and {\it GALEX} FUV and
NUV.  Average FUV, NUV and 24\,$\micron$ luminosities are obtained for
each subset of galaxies.  Optical photometry in the $U, B, V, R$ and
$I$ bands from the COMBO-17 survey is available for all individual
sample galaxies, allowing calculation of the average luminosities 
in these bands.

The total UV luminosity is estimated by integrating the spectral 
energy distributions over the
wavelength range of rest-frame 1500 - 2800\,\AA\, from linear
interpolation of the FUV, NUV and optical $U, B, V, R$ and $I$ bands.
We estimate the total IR luminosity from the observed 24\,$\micron$
luminosities using three sets of luminosity-dependent IR SED templates
\citep{Lagache04,Dale02,Chary01}.
\footnote{Adoption of the updated IR SED templates based on 
Spitzer observations \citep{Rieke09} gives consistent results 
within the uncertainties. }
Although these templates are
derived from local star-forming galaxies, they can be used to
represent IR SEDs for distant star-forming galaxies
\citep{Marcillac06,Zheng07,Magnelli09}. 
We calculate the SFR from the UV and IR luminosities assuming 
a Kroupa IMF, following \citet{Bell05}.  Average SFR values are 
therefore obtained for each subset of galaxies split by redshift in three
samples.  We combine bootstrap errors with scatter between templates
to compute errors in estimating the SFR.  Systematic errors in estimating SFR
have not been included explicitly, but are discussed in
\citet{Bell05} and \citet{Zheng06}. 
We also compute the average stellar mass for
each subset of sample galaxies using stellar mass estimates obtained
from COMBO-17 optical SEDs by \citet{Borch06}. Table~\ref{results}
lists the results, including object number, average SFR, average
stellar mass and volume-averaged SFR for our samples.

\begin{deluxetable}{llccc}
\tabletypesize{}
\tablewidth{0pt}
\tablecaption{Average SFR and Stellar mass for three galaxy sub-populations
  over $0<z<1$.}
\tablehead{   z  & N$_{\rm obj}$ & 
    $<SFR>$  & $<M_\ast>$  & $\rho_{\rm SFR}$       \\
                 &               & 
($\sfrunit$) & (10$^{10}$\,M$_\odot$) & ($10^{-2}\,\sfrunit$\,Mpc$^{-3}$)}
\startdata
\multicolumn{5}{c}{massive galaxies} \\
0.04  & ... & ... & ... & 0.4$\pm$0.2 \\
0.3 & 64  & 3.1$\pm$0.2  & 5.0$\pm$0.5 & 1.1$\pm$0.5 \\
0.5 & 230 & 4.3$\pm$0.2  & 4.5$\pm$0.6 & 1.2$\pm$0.4 \\
0.7 & 421 & 8.3$\pm$0.3  & 5.7$\pm$0.5 & 2.5$\pm$0.7 \\
0.9 & 188 & 15.1$\pm$0.8 & 5.5$\pm$0.7 & 3.5$\pm$1.9 \\
\multicolumn{5}{c}{massive spheroid-dominated galaxies} \\
0.04 & ... & ... & ... & ... \\
0.3 & 35  & 0.9$\pm$0.1 & 4.5$\pm$0.8 & 0.22$\pm$0.12 \\
0.5 & 122 & 1.3$\pm$0.1 & 6.4$\pm$0.8 & 0.22$\pm$0.07 \\
0.7 & 249 & 4.2$\pm$0.2 & 6.7$\pm$0.7 & 0.65$\pm$0.19 \\
0.9 & 87  & 9.1$\pm$0.8 & 6.8$\pm$1.2 & 1.01$\pm$0.59 \\
\multicolumn{5}{c}{massive $SFR\ge 3\sfrunit$ galaxies} \\
0.04 & ... & ... & ... & 0.2$\pm$0.1 \\
0.3 & 19  & 7.1  & 4.1 & 0.7$\pm$0.4 \\
0.5 & 89  &10.2  & 4.9 & 1.1$\pm$0.6 \\
0.7$^\mathrm{a}$ & 176 & 15.2 & 4.5 & $>$1.9$\pm$0.8 \\
0.9$^\mathrm{a}$ & 81  & 23.0 & 4.6 & $>$2.3$\pm$1.4 \\
\multicolumn{5}{c}{massive $SFR\ge 10\sfrunit$ galaxies} \\
0.04 & ... & ... & ... & 0.04$\pm$0.03 \\
0.3 & 3   & 14.3 & 5.9 & 0.2$\pm$0.1 \\
0.5 & 29  & 20.2 & 4.9 & 0.7$\pm$0.2 \\
0.7 & 99  & 22.0 & 4.8 & 1.6$\pm$0.5 \\
0.9 & 63  & 27.4 & 4.0 & 2.1$\pm$1.2 \\
\enddata
\tablenotetext{a}{Sample is incomplete in this redshift bin. The average SFR
  is overestimated and the volume-averaged SFR is underestimated.}
\label{results}
\end{deluxetable}

We further calculate the SFR for sample galaxies that are individually detected
at 24\,$\micron$. The IR luminosities of these galaxies typically 
dominates ($>80$\%) the bolometric luminosity (UV+IR).  
The 5\,$\sigma$ detection limit of
our 24\,$\micron$ imaging is 83\,$\mu$Jy. This limit corresponds to an
IR luminosity $\sim 3 (10)\times 10^{10}$\,L$_\odot$ and a SFR
roughly $\sim3 (10) \sfrunit$ at $z=0.6$ (1).  Our sample of massive
galaxies with $SFR\ge 3\sfrunit$ is therefore incomplete at
$z>0.6$; a cut of $SFR \ge 10\sfrunit$ is complete up to $z=1$.
We bear this in mind and discuss its effects on the relevant conclusions.

We could simply use the SFR density determined from the ECDFS 
alone in what follows, by simply dividing the total 
amount of SFR in these subsamples by the survey volume.
That approach would give very similar results to those that we present
later (with the exception of the $0.6<z<0.8$ bin, where a somewhat 
higher SFR density would be determined owing to the overdensity in 
the ECDFS at $0.6<z<0.8$), and our conclusions would be unchanged.
Yet, we choose to use a slightly more complex approach that attempts
to reduce field-to-field variance by comparing the ECDFS to the rest
of the COMBO-17 survey.  
We adjust the SFR densities for the massive galaxy 
samples by multiplying by the ratio of the average stellar mass density derived for all of COMBO-17 to the stellar mass density in massive galaxies
in the ECDFS:
i.e., $SFR_{\rm all} \sim SFR_{\rm ECDFS} \times \rho_{\ast,\rm all} / \rho_{\ast,\rm ECDFS}$.  Similarly, 
we adjust the inferred SFR density in spheroid-dominated 
galaxies using the stellar mass density in red-sequence galaxies
from \citet{Borch06} as a guide \citep[see][for the comparison between
mass functions of morphology-selected and color-selected early-type
galaxies]{McIntosh05}: $SFR_{n>2.5,\rm all} \sim SFR_{n>2.5,\rm ECDFS} \times \rho_{\ast,\rm red,all} / \rho_{\ast,\rm red,ECDFS}$.  

\begin{figure*} \centering
  \includegraphics[width=0.85\textwidth]{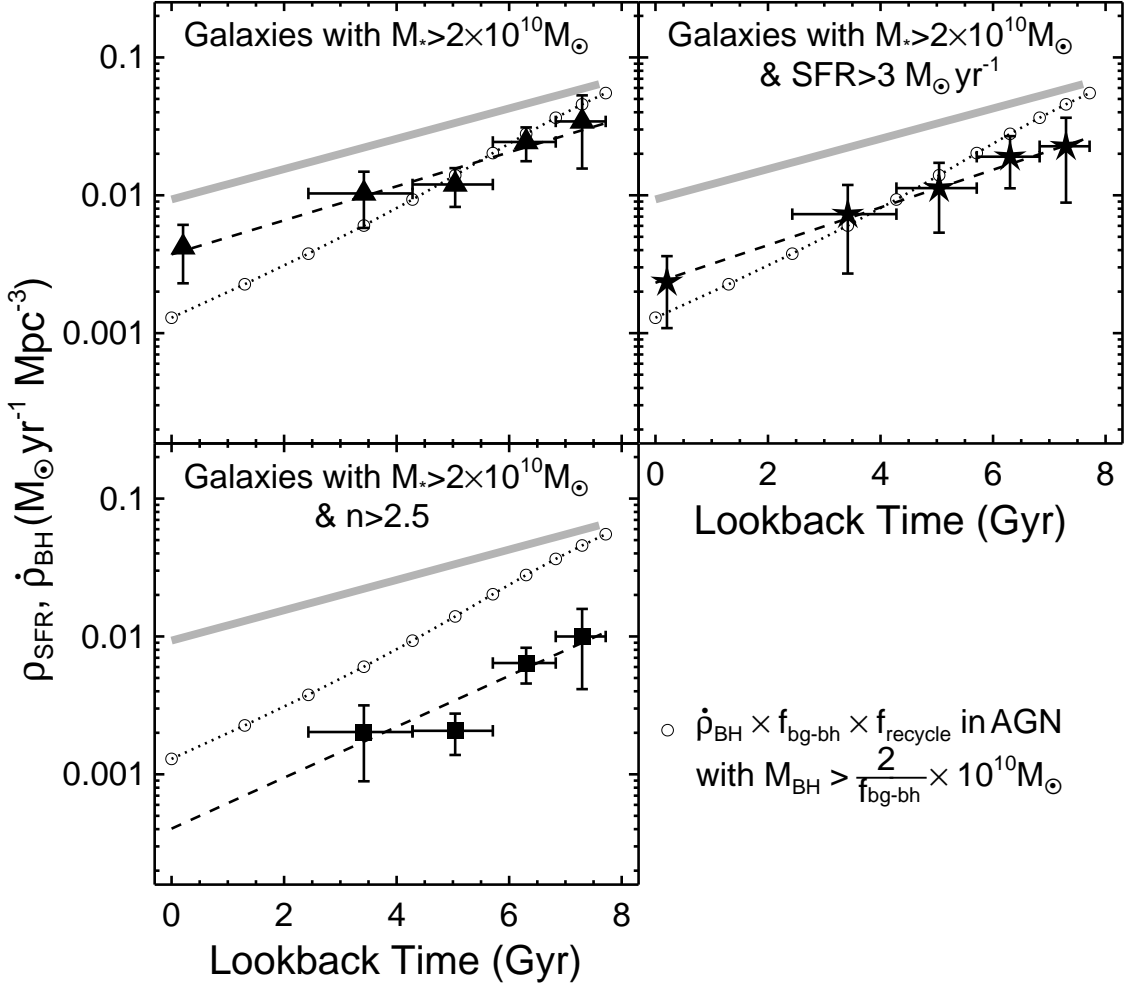}
\caption{The volume-averaged SFR as a function of lookback time for
  massive ($M_\ast \ge 2\times10^{10}$\,M$_\odot$) galaxies ({\it
  top-left}), massive high-SFR ($SFR\ge 3$\,M$_\odot$\,yr$^{-1}$)
  galaxies ({\it top-right}) and massive spheroid-dominated (S\'ersic
  index $n> 2.5$) galaxies ({\it bottom-left}).  Horizontal errorbars
  represent the redshift range and vertical errorbars represent
  1\,$\sigma$ errors derived from bootstrapping.  The dashed lines show
  least-squares fits to data points (taking into account the errorbars
  on both axes).  
  The gray thick line in each panel is the least-squares fit to 
  all data points at $z<1$ listed in Figure~\ref{sfr}, giving 
  best estimates of the global volume-averaged SFR.
  The dotted lines are the same in all three
  panels, showing the volume-averaged BHAR attributed to
  luminous AGN ($L_{\rm bol}\ge 10^{44.9} $erg\,s$^{-1}$) converted
  from the best-fit LDDE model of the bolometric LFs given in
  HRH07 with a radiation efficiency $\epsilon=0.1$, 
  scaled by a factor of $f_{\rm bg-bh}\times f_{\rm recycle}$. 
  Here $f_{\rm bg-bh}$ is the local $M_{\rm BH}/M_\ast$ ratio of 650 
  given by the local BH-bulge mass relation
  \citep{Haring04}, and $f _{\rm recycle}$ is the recycling factor. We
  adopt $f _{\rm recycle}=2$.  Assuming an average Eddington ratio
  $\log$ ($L_{\rm bol}/L_{\rm Edd}) = -0.60 \pm$0.3 for luminous AGN
  \citep{Kollmeier06}, the bolometric luminosity $L_{\rm bol}\ge
  10^{44.9}$erg\,s$^{-1}$ corresponds to BH mass $M_{\rm BH} \ge
  2.6\times 10^7$\,M$_\odot$ and accordingly spheroid mass $2\times
  10^{10}$\,M$_\odot$.}
\label{coevolution}
\end{figure*}
\subsubsection{Local comparison sample}

We augment the intermediate-redshift sample with a 
sample of 2177 local galaxies collected from the NASA/IPAC
Extragalactic Database (NED) to assess the corresponding
volume-averaged SFR at $z\sim 0$ for the three sub-populations.  The
sample galaxies are selected with a 2MASS $K$-band magnitude cut $K<12$
in the volume of $1500 \leq cz$ (km\,s$^{-1}) \leq 3000$ and Galactic
latitude $b > 30\degr$. About 61\% of the selected galaxies have
redshifts from NED.  We believe that the redshift identification is
not significantly biased and the sample is representative of local
galaxies.  Details about the sample completeness can be found in
\citet{Bell05}. Stellar masses were estimated from the $K$-band
absolute magnitude assuming a $K$-band stellar mass-to-light ratio of
0.6\,M$_\odot$/L$_\odot$ and a Kroupa IMF \citep{Bell03}.  Of the 2177
sample galaxies, 1089 have IRAS 60 and 100\,$\micron$ detections. 
The IRAS detection limit
of $\sim 3 \times 10^{-11}$\,ergs\,cm$^{-2}$\,s$^{-1}$ is applied to
the remaining 1088 galaxies.  The total IR luminosity derived from
IRAS observations is used to estimate the SFR following \citet{Bell03a}.
The typical error is $\sim 0.3$\,dex for both stellar mass and SFR.
We calculated the volume-averaged SFR for the mass limited subsample,
and for the mass and SFR limited subsample. The morphological S\'ersic
index parameter is not available for this local sample, so the
volume-averaged SFR at $z=0$ is missing for the mass and morphology
limited subsample. We note that the local comparison sample 
provides roughly consistent results with the optically-selected 
sample from the SDSS \citep{Schiminovich07}, although a small offset 
exists between the two mainly due to the difference in SFR estimator.

\subsubsection{Results}\label{sfr_res}

Figure~\ref{coevolution} shows the volume-averaged SFR as a function
of the lookback time for three galaxy subsamples.  For
comparison, we also show the volume-averaged SFR density for all galaxies (\S
\ref{sfrhis}; thick gray line).  All three galaxy subsamples 
have a higher average SFR at larger redshift, consistent with the overall galaxy
population.  A straight line is fit to the data points, accounting
for the error bars on both axes.\footnote{The vertical error 
shows the random uncertainties in SFR density. 
The horizontal ``errorbar'' indicates the
redshift range and the data points are uniformly weighted in redshift
space.}  The line is described by $\log \rho_{\rm SFR} = {\rm A}
\,t_{\rm LB} - {\rm B}$, where $\rho_{\rm SFR}$ is the volume-averaged
SFR and $t_{\rm LB}$ is the lookback time in units of Gyr. The
best-fit parameters given by the error-weighted least-squares fit are
listed in table~\ref{fitpara}.

\begin{deluxetable}{lccl}
\tabletypesize{}
\tablewidth{0pt}
\tablecaption{Best-fit parameters for the increases of volume-averaged SFR/BHAR of galaxy/AGN sub-populations with lookback time.}
\tablehead{population & A$^\mathrm{a}$  & B$^\mathrm{a}$ & Ref.$^\mathrm{b}$}
\startdata
all galaxies           & 0.11 $\pm$ 0.01 & -2.03 $\pm$ 0.03 & this work \\
massive$^\mathrm{c}$   & 0.12 $\pm$ 0.05 & -2.43 $\pm$ 0.26 & this work \\
massive \& $n>2.5$     & 0.18 $\pm$ 0.11 & -3.40 $\pm$ 0.61 & this work \\
massive \& $SFR\ge $3  & 0.14 $\pm$ 0.06 & -2.63 $\pm$ 0.28 & this work \\
massive \& $SFR\ge $10 & 0.25 $\pm$ 0.06 & -3.46 $\pm$ 0.31 & this work \\
AGN ($\log L_{\rm bol}>$43)   & 0.17 & -2.24 & H07,M04 \\
AGN ($\log L_{\rm bol}>$44.9) & 0.21 & -2.73 & H07,M04 \\
AGN ($\log L_{\rm bol}>$45.5) & 0.23 & -3.03 & H07,M04 \\
\enddata
\tablenotetext{a}{Best fit parameters for $\log \rho_{\rm SFR}={\rm A}t_{\rm LB}+{\rm B}$ (galaxies) or $\log \dot{\rho}_{\rm BH}+3.11={\rm A}t_{\rm LB}+{\rm B}$ (AGN) 
   over the redshift range $0<z<1$, where $t_{\rm LB}$ is the lookback time in
   units of Gyr. The constant 3.11 is the scaling factor 1300 in logarithm.}
\tablenotetext{b}{References --- B07: \citet{Bell07}; LF05: \citet{LeFloch05};
  H07: \citet{Hopkins07}; M04: \citet{Marconi04}; } 
\tablenotetext{c}{Refer to $M_\ast\ge 2\times 10^{10}$\,M$_\odot$.}
\label{fitpara}
\end{deluxetable}

Integration of the volume-averaged SFR
over $0<z<1$ shows that massive galaxies contribute 45\% to the
overall volume-averaged SFR.  Most of this SFR ($>$85\%) is 
contained in the mass-limited subsample defined to have 
SFR $> 3 M_{\sun}$\,yr$^{-1}$; this reflects that the typical
SFR of a massive star-forming galaxy is $\ga 3M_{\sun}$\,yr$^{-1}$ 
\citep{Noeske07}.  A subsample limited at $> 10 M_{\sun}$\,yr$^{-1}$
contains 43\% of the SFR in massive galaxies.

Star formation in the host galaxies of X-ray-detected AGN is not counted. 
The difficulty in estimating star formation in these galaxies is how to
separate the IR emission powered by star formation from that by AGN. 
We note that the removal of X-ray detected sources in our sample
selection does not influence our results.  Indeed, the 24\,$\micron$
fluxes from the X-ray detected sources are negligible compared with
the total 24\,$\micron$ fluxes from those X-ray undetected ones
($<15$\%; see also \citealt{Brand06}).  The average SFR is dominated
by galaxies without significant AGN, indicating that SF and BH
accretion take place in different phases or over different time
intervals.

\subsubsection{Hosts of star formation: spheroids or disks?}\label{SFhost}

About one quarter of the SF in massive galaxies is contained
in `spheroid-dominated' galaxies with $n>2.5$; given that 45\% of
the total SFR density at $z \la 1$ is contained in the massive
galaxy sample, this result is consistent with earlier results 
that found that $<10$\% of all SF is in lenticular or elliptical 
galaxies \citep{Bell05}.  

Our SFR estimates measure the SF in the whole galaxy.  The IR data we
use do not have sufficient spatial resolution to distinguish between
SF in spheroidal and disk components within a galaxy.
Accordingly, we adopted a simplistic approach of 
assigning all SF in concentrated $n>2.5$ spheroid-dominated galaxies 
to spheroids (an overestimate), and assigning all SF in $n<2.5$ disk-dominated
systems to disks (some of that SF may be in spheroids in these disk-dominated 
galaxies; this would drive one towards underestimating the SFR
in disks).  While mid-IR imaging data from JWST will help to address 
this problem in the long-term, our present approach is 
essentially all that the current data support, and provides
a reasonable guide to the possible contribution of spheroids to 
the volume-averaged SFR density.  Bearing in mind the 
caveats, we conclude nonetheless that the bulk of the
overall SF is associated with disks, rather than spheroids; we
estimate that $<30$\% of SF in massive galaxies can be associated
with $n>2.5$ spheroid-dominated galaxies at any epoch in the last 
8 Gyr. 

\subsection{BH accretion in Luminous AGN}\label{bhacc}

It would be ideal to measure the BH accretion rate for our sample of 
galaxies in the ECDFS.  Unfortunately, episodes of intense accretion 
by supermassive black holes are rare, and the number of AGN in 
the ECDFS is too low to allow an analysis as detailed as our above 
discussion of the SFR density.  Accordingly, we here adopt 
an intermediate approach based on luminosity-selected
samples (derived from larger-area surveys) that gives a good
estimate of the BHAR density in the massive galaxy population.

It is generally assumed that BH accretion obeys the Eddington limit.
A luminous AGN is normally caused by intense accretion onto a massive
BH, while a faint AGN can be due to either a massive BH with low
accretion rate or a less massive BH with higher accretion rate.
\citet{Kollmeier06} found that luminous AGN have an average Eddington
ratio $L_{\rm bol}/L_{\rm edd}$ $\simeq 0.25$ with a scatter of
0.3\,dex over a wide redshift range.  This implies a strong
correlation between AGN luminosity and BH mass.

We want to estimate the BHAR in massive BHs in galaxy populations 
of mass $M_\ast\ge 2\times10^{10}$\,M$_\odot$, and compare it to the SFR
in the same galaxies.
According to the local BH-bulge mass relation given by
\citet{Haring04}, spheroids with mass $M_\ast\ge
2\times10^{10}$\,M$_\odot$ host SMBHs with mass $M_{\rm BH}\ge
2.6\times 10^{7}$\,M$_\odot$.  Combined with the local mass function
from \citet{Bell03}, we assume a mean spheroid-BH mass ratio
$M_\ast/M_{\rm BH} = 650$ for local massive spheroid-dominated
galaxies.  Assuming Eddington ratio $L_{\rm bol}/L_{\rm edd} =0.25$ and
radiation efficiency $\epsilon =0.1$, we infer that massive BHs of
$M_{\rm BH}\ge 2.6\times10^{7}$\,M$_\odot$ power luminous AGN of $\log
L_{\rm bol}\ge 44.9$ with $ BHAR \geq 0.13 \sfrunit$.  We have seen
that such intense BH accretion events statistically match starburst
events with SFR $\geq 3\sfrunit$, according to the universal agreement
between BH accretion events and SF events (a factor of 20 in
intensity; see \S\ref{sec:multifuncs}).
%


If the local BH-bulge mass relation holds at all cosmic
epochs, then the mass growth of SMBHs and that of spheroids should
follow the same relation as $\frac{<\dot{M}_\ast>}{<\dot{M}_{\rm BH}>}
= 650$. We use the best-fit LDDE model of bolometric LFs from
HRH07 to calculate the volume-averaged BHAR contained in
luminous AGN of $\log L_{\rm bol}\ge 44.9$ over $0<z<1$.  The results
are scaled up by a factor of 1300 and shown with the dotted lines in
Figure~\ref{coevolution} (same in all four panels).  The scaling
factor 1300 accounts for the mean local spheroid-BH mass ratio 650 and
the recycling factor of two. 
 Note that some studies \citep[e.g.][]{Treu07} have claimed to
  find evidence that the BH-spheroid mass ratio was larger in the past
  (although see \citet{Lauer07b}). If this is the case, it would
  obviously invalidate the ``strong'' co-evolution picture. It would
  impact our calculation by changing the slope of the BHAR function
  when cut by the ``matching'' BH mass, because a given bulge mass
  would correspond to a larger BH mass and therefore to a higher AGN
  luminosity at high redshift. We discuss the implications of such an
  effect in Section~\ref{sec:summ}. 

The open circles mark redshifts from $z=0$ to $z=1$ with a step size
$\Delta z=0.1$. We fit a straight line to these points and present the
best-fit parameters given by the least-squares fit in
Table~\ref{fitpara}. In addition, we also fit the overall BHAR (in AGN
of $\log L_{\rm bol}>43$) and tabulate the corresponding best-fit
parameters in Table~\ref{fitpara}.  We estimate that luminous ($\log
L_{\rm bol}\ge 44.9$) and very luminous ($\log L_{\rm bol}\ge 45.5$)
AGN are responsible for $\sim 70$\% and $\sim 45$\% of the overall BH
accretion over 0$<z<$1, respectively.  We caution that the
correspondence between AGN luminosity and BH mass depends on the
Eddington ratio.  The uncertainty of 0.3\,dex in the Eddington ratio
\citep{Kollmeier06} causes an negligible error in the sum of the
intense BH accretion. The scatter in the local BH-bulge relation
($\sim 0.3$\,dex) introduces a comparable error.  
It is worthwhile to note that the Eddington ratio may be luminosity
  dependent: it is possibly close to unity for luminous quasars and
  increasingly sub-Eddington for low-luminosity AGN
  \citep{Babic07,Bundy08}. If so, our calculations will underestimate
  the accretion for massive BHs. However, as long as the {\it
    average} accretion rate for the objects we include in our
  calculation is similar to that of the Kollmeier et al. sample, our
  results should be reasonably accurate.
In any case, the overall BH accretion is almost certainly dominated
by luminous AGN associated with massive BHs.
We emphasize that our conclusions essentially rely on the relative 
density, i.e., the BHAR density of luminous AGN to the total, and
are marginally affected by how the numbers are calibrated.

\subsection{Host galaxy connections} \label{sect:host}

We showed in \S \ref{sfr_res} that $\sim$45\% of SF happens in massive
galaxies ($M_\ast\ge 2\times 10^{10}$\,M$_\odot$).  We estimated in \S
\ref{bhacc} the total amount of BH accretion in massive galaxies using
an accretion rate-limited sample, in conjunction with an assumption
about the distribution of Eddington ratios, to conclude that $\sim
70$\% of BH accretion should be contained in the massive galaxy
population.  The uncertainties inherent in our analysis in \S
\ref{bhacc} are considerable, but it appears that the bulk of BH
accretion tends to occur in massive systems, whereas much of the SF
happens in lower mass systems.

Furthermore, we showed in \S \ref{SFhost} that $<30\%$ of the star
formation in massive galaxies ($M_\ast\ge 2\times 10^{10}$\,M$_\odot$)
happens in spheroid-dominated systems: star formation, to first order,
happens in disks.  
Although a substantial fraction of AGN by
  number reside in late- and intermediate-type galaxies
  \citep{Pierce07,Gabor09}, the luminous ones tend to reside in
  massive, early-type galaxies. 
Therefore, it is clear that the bulk
of the black hole accretion happens in spheroid-dominated galaxies: it
appears that BH accretion is happening in different objects or at a
different phase in the life-cycle of galaxies than the bulk of the
star formation.

\begin{figure} \centering
  \includegraphics[width=0.48\textwidth]{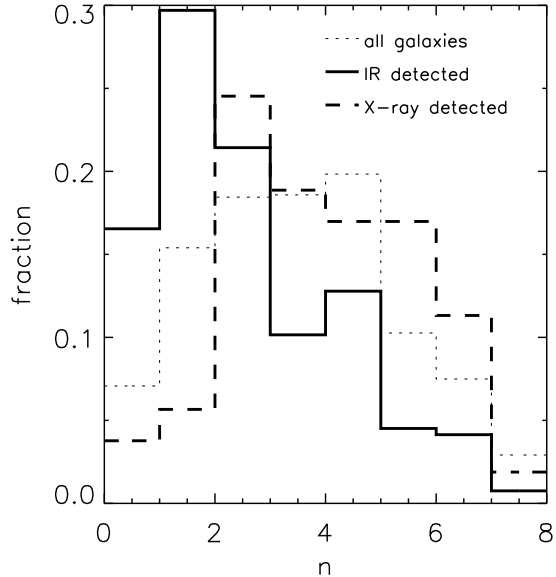}
\caption{Histograms of S\'ersic index $n$ for IR-detected galaxies
(the {\it thick-solid} line) and X-ray-detected galaxies (the {\it
thick-dashed} line) for galaxies with $M_\ast > 2\times
10^{10}$\,M$_\odot$ and $0.4<z<0.8$.
Here the IR-detected galaxies refer to star-forming galaxies detected
at 24\,$\micron$ and with soft X-ray luminosity
$<10^{42}$\,erg\,s$^{-1}$.  The X-ray-detected galaxies are those
detected in the {\it Chandra} soft band with X-ray luminosity
$>10^{42}$\,erg\,s$^{-1}$, suggestive of AGN activity.  The
X-ray-detected AGN host galaxies are more concentrated than the
IR-detected star-forming galaxies.
 } \label{agn}
\end{figure}

We illustrate this point with an example from the ECDFS (this exercise
simply reinforces the conclusions of, e.g.,
\citealp{Grogin05,Nandra07,Pierce07,Georgakakis08,Alonso08,Gabor09}).
We select massive galaxies with $M_\ast > 2.5\times
10^{10}$\,M$_\odot$ in the redshift range $0.4<z<0.8$ in the ECDFS to
create a parent sample. From the sample, we select those detected in
the 250\,ks {\it Chandra} X-ray observation \citep{Lehmer05} to make a
subsample of galaxies having luminous AGN with soft X-ray luminosity
$>10^{42}$\,erg\,s$^{-1}$.  The IR-detected galaxies, i.e. those
detected at 24\,$\micron$, are selected to comprise a subsample of
star-forming galaxies \citep[see \S\ref{essfr} for the details of the
  MIPS 24\,$\micron$ observation and][for further
  discussion]{Donley08}.  This subsample is also limited to have soft
X-ray luminosities $<10^{42}$\,erg\,s$^{-1}$ to avoid AGN
contamination to the IR.  S\'ersic indices are derived for all sample
galaxies from their GEMS/ACS images \citep{Haussler07}.  The
X-ray-detected AGN are generally obscured in the optical and
contribute marginal contamination ($< \sim 10$\% of central flux) to
their host galaxies. Therefore the S\'ersic index is derived from the
GEMS/ACS imaging of the host galaxies is insensitive to the central
AGN.  Figure~\ref{agn} shows the histograms of S\'ersic index $n$ for
all galaxies, subsamples of IR-detected galaxies and X-ray-detected
galaxies. One can see that two thirds of X-ray-detected galaxies have
$n > 3$, while two thirds of IR-detected galaxies have $n <$3,
suggesting that the X-ray-detected AGN host galaxies are more
concentrated than the IR-detected star-forming galaxies.  A
Kolmogorov-Smirnov (K-S) test indicates that the probability for both
distributions to be drawn from the same distribution is less than
10$^{-9}$, implying that the IR-detected galaxies and X-ray-detected
galaxies are different populations.  Put simply, star formation and
SMBH accretion take place in dramatically different systems:
the majority of the global star formation happens in late-type galaxies
  (i.e. disks), while the dominant phases of BH accretion occur in
  intermediate- and early-type galaxies.

\section{Summary and Discussion}
\label{sec:summ}

We have shown that the global SF history is proportional to the global
BH accretion up to (at least) $z\sim 1$, and the proportionality
factor (2000) is that expected if the BHAR is a fixed fraction of the
star formation that ends up in galactic bulges. This parallel between
the cosmic SF history and the cosmic BH accretion history is an
important constraint on the co-evolution between SMBHs and
galaxies. It suggests that {\it overall} accretion-driven BH growth
and {\it overall} SF-driven galaxy growth trace each other, and that
the globally averaged BH mass to stellar mass ratio remains
constant. Furthermore, we find that the SFR and BHAR {\it multiplicity
  functions} are also scaled versions of one another at $z<1$,
suggesting that the intensity of star formation and BH accretion are
at least statistically linked.

We then refined our analysis to test the hypothesis that star
formation and BH accretion do not just trace each other in a
statistical fashion, but that {\it the SFR and BHAR are related by a
fixed factor} $f_{\rm co} \equiv f_{\rm recycle} \, f_{\rm bg-BH} \simeq
1300$ {\it in every bulge/BH-building event}. We focused our
analysis on massive galaxies with $M_* > 2 \times
10^{10}$\,M$_{\sun}$, for which our observational sample is complete
to $z\sim 1$. Assuming that the local relationship between BH mass and
bulge mass remains constant, spheroids with $M_* > 2 \times
10^{10}$\,M$_{\sun}$ should host BH with mass $M_{\rm BH}\ge 2.6\times
10^{7}$\,M$_\odot$. Assuming an average Eddington ratio $L_{\rm
bol}/L_{\rm Edd}=0.25$ \citep{Kollmeier06}, and $\epsilon =0.1$, these
BH, when active, power AGN with $BHAR \geq 0.13 \sfrunit$ or
bolometric luminosity $\log L_{\rm bol}\ge 44.9$. We computed the BHAR
contributed by AGN with bolometric luminosities above this limit, and
compared it with the SFR contributed by galaxies with $M_* > 2 \times
10^{10}$\,M$_{\sun}$. We found that the SFR and BHAR again trace each
other  but are offset by approximately the factor of $f_{\rm
co} \simeq 1300$ expected for massive spheroid-dominated galaxies.
This result is nearly unchanged if we compute the SFR from galaxies 
limited in both stellar mass and SFR (SFR\,$\ge 3\sfrunit$), 
as these ``high-intensity'' SF events dominate the SFR
budget in massive galaxies. 
The massive galaxies are simply dominating the total SFR, and the big
accretion systems are dominating the accretion history (although with a
little more evolution). Just as before when we compared 
the overall SFRD with the overall BHAR we got a match, 
we should get a match for the top two panels.

Further examination of the nature of the galaxies hosting star formation
and BH activity showed that only about 30 \% of the SF ``budget'' 
needed to match the BHAR is observed in massive {\it spheroid dominated}
galaxies. Furthermore, the distribution of S\'ersic indices for star
forming galaxies and that for AGN detected in the soft X-ray indicate
that these two classes of activity tend not to reside in the same types of host
galaxies. The bulk of star formation occurs in {\it isolated disks}
while the majority of luminous AGN hosts are intermediate- and 
early-type galaxies
\citep[see also][]{Grogin05,Nandra07,Georgakakis08,Alonso08,Watson09,Gabor09}. 
These results clearly rule out the simplest picture of object-by-object
co-evolution, in which BH and bulges grow proportionally and simultaneously.
This is also supported by the detection of a small fraction of luminous 
X-ray detected AGN in spirals, indicating that BH accretion may happen in 
non-bulge-building events \citep[e.g.,][]{Georgakakis09}.

 There are both observational claims and theoretical expectations
  that the relationship between BH mass and spheroid mass might have
  been larger in the past. While this evolution would impact our
  calculation by shifting the mass of the BH (and therefore the cutoff
  luminosity for AGN) to include in our ``matched'' BHAR, this would
  only cause a minor change in the slope of the BHAR and cannot remove
  the mismatch between SF and BHA in spheroid dominated
  galaxies. Therefore, this would not change any of our conclusions.

There are two possible ways to reconcile this apparent paradox.  One
possibility (which shall now call Merger Scenario 1) is that bulges
and BH grow in proportion to one another, and in the same {\it
  events}, but with an offset in time.  For example, in simulations of
galaxy-galaxy mergers including BH growth, the luminous AGN phase is
delayed by about a dynamical time (a few hundred Myr) relative to the
peak of the starburst activity \citep{DiMatteo05,Hopkins05,Hopkins06}.
In this picture, we would expect that most AGN hosts would resemble
late-stage mergers, i.e. objects with spheroidal morphology but young
stellar populations, in plausible agreement with our observational
results. Moreover, in this picture, we would expect that a small fraction
of AGN to be detected in early stage mergers, and the star
formation associated with this BH growth should be observed in
early-stage mergers, i.e. close pairs or highly morphologically
disturbed objects.

However, other studies have found that only a small fraction of the
global star formation at $z<1$ is occurring in galaxy mergers
\citep{Bell05,Wolf05,Jogee09,Robaina09}: the strongest limit
  is set by Robaina et al., who find that $<$10\% of the star
  formation in massive galaxies ($m_* > 10^{10} M_\odot$) is directly
  triggered by {\it all identifiable phases} (from close pairs to
  morphologically disturbed remnants) of major galaxy merging.
This is strong evidence against the Merger Scenario 1.
  Therefore, while a small fraction of the star formation associated
  with the BH growth that we witness at $z\la 1$ may be occurring in
  the same merger events that feed the BH, it seems that this does not
  represent the majority of the SF activity needed to ``match'' the
  observed BH accretion.

The second possibility (Merger Scenario 2) is that most star formation
takes place in isolated disks. When these disks merge, the
pre-existing stars are scrambled into a dynamically hot spheroidal
remnant\footnote{In what follows, much of what we argue for merger scenario 2 would also apply if the main route for forming low-mass spheroids were secular evolution.  The key point for this paper is that the bulk of the stars that end up in the spheroid are formed long before the epoch of actual bulge creation, which applies equally in Merger Scenario 2 and secular evolution.}. The presence of even a modest amount of gas leads to the
formation of a remnant with higher phase-space density and hence a
deeper gravitational potential well
\citep{Dekel-cox06,Robertson06,Cox06}. If the depth of the potential
well in the central parts of the galaxy determines how large the BH
can grow, as in the picture of self-regulated BH growth outlined by
\citet{Hopkins07a}, then the post-merger BH will end up on the BH
mass-bulge mass relation, {\it despite the fact that the SFR and BHAR
  during the merger event may not necessarily obey the universal
  proportionality factor}.

  One can easily see that in this scenario, in many cases the BH
  accretion and the star formation will not trace each other during
  the merger event. For example, in a major merger of two pure disk
  progenitors, the scaled BHAR will be larger than the SFR associated
  with bulge growth. Most of the bulge mass will be contributed by the
  pre-existing stars in the progenitor disks, and the BH will grow to
  ``catch up'' with the bulge. In the opposite extreme, in a merger of
  two galaxies with pre-existing massive BH, there may be very little
  BH accretion because the AGN will almost immediately ``shut itself
  off''. In this case, the SF in the merger might be larger than the
  scaled BH accretion (if the progenitors contain enough gas). 

  Because what we see is more BH accretion than we would expect
  based on the amount of star formation in spheroids, we conclude that
  most of the mass in these spheroids was contributed by the
  pre-existing stars in the progenitors rather than by new stars
  formed in the merger-associated bursts.  
Indeed, we have independent lines of evidence that this must be the
case. We know that the colors of elliptical galaxies and bulges at
$z\la 1$ are characteristic of fairly old stellar populations
\citep[][and references therein]{Gallazzi06,Renzini06}, and similarly,
fossil evidence from line strengths in nearby ellipticals also
precludes the very recent formation of a large fraction of the stellar
mass \citep{Trager00,Thomas05}. 
  In addition, this picture is consistent with the relatively small
  fraction of star formation associated with ongoing mergers at $z<1$:
  in cosmological models that incorporate a scenario like Merger
  Scenario 2 \citep[e.g.][]{Somerville08}, the predicted fraction of
  star formation triggered by major mergers at $z<1$ is about 7\%, in
  excellent agreement with the observational estimates (see Robaina et
  al. 2009 for details of the comparison). 

  A slightly modified version of Merger Scenario 2 can also be
  accommodated in a Universe in which the BH-spheroid mass ratio has
  decreased with time. This is in fact expected in the theoretical
  picture that has emerged from the analysis of hydrodynamic
  simulations of galaxy mergers with self-regulated BH growth
  \citep[e.g.][]{Robertson06b,Hopkins07a}. These studies find that
  the final BH-spheroid mass ratio depends on the gas fraction of the
  progenitors. This is because gas-rich progenitors can dissipate
  energy during the merger, and all else equal, form more compact
  remnants than gas-poor mergers
  \citep{Dekel-cox06,Robertson06,Cox06,Covington08}. In this deep
  potential well, the BH can grow to a larger mass (relative to the
  spheroid) before the pressure-driven outflow halts further
  accretion.  If gas fractions in merging galaxies were higher at high
  redshift, as is generally expected, this would then lead to the
  prediction that BH were more massive relative to their spheroids at
  high redshift. A significant fraction of mergers at $z<1$ are
  expected to involve at least one galaxy that already contains a
  massive BH. If the pre-existing BH is already larger than the
  ``critical mass'' set by the new remnant, further BH growth will be
  stifled. Theoretical models predict that this expected evolution is
  relatively mild: less than a factor of two since $z\sim1$
  \citep{Hopkins07a}.

\citet{Borch06} have shown that the stellar mass in red (predominantly
spheroidal) galaxies has increased by a factor of 2--3 since $z\sim1$,
while the stellar mass in blue (disk-dominated) galaxies has remained
about the same over this time period. At the same time, we know that
the bulk of new star formation is occurring in blue (disk-dominated)
galaxies, implying that mass in blue (disk) galaxies must be
transformed into red (spheroidal) galaxies at approximately the same
rate that new stars are forming \citep{Bell07}. 
  Given that most
  of the red galaxies are spheroid-dominated, which is well
  established at least at $z<1$ \citep{Bell04b}, the mass on the red
  sequence cannot grow simply via the quenching of star formation in
  disks --- a strong dynamical process like merging is required.

Taken together, this evidence strongly favors a picture in which, at
least at $z<1$, the bulk of the growth of mass in spheroids is due to
merging of pre-existing, already old disk stars, and BHs
preferentially grow to ``catch up'' to their newly assembled bulges
following a merger event.  In this case, the parallel evolution of the
global SFR and BHAR may simply reflect the apparent coincidence that,
at $z<1$, the rate of formation of new stars is approximately equal to
the rate that stellar mass is transferred from disks into bulges via
mergers.

\acknowledgments

We thank the referee for valuable comments that improved this manuscript. 
X.\ Z.\ Z.\ is supported by NSFC under grant 10773030,10833006 and by 
the National Basic Research Program of China (973 program; 2007CB815404). 
E.\ F.\ B.\ and K.\ J.\ thank the Deutsche Forschungsgemeinschaft for 
their support through the Emmy Noether Program.  
This research has made use of the NASA/IPAC Extragalactic Database (NED)
which is operated by the Jet Propulsion Laboratory, California
Institute of Technology, under contract with the National Aeronautics
and Space Administration.

\end{document}